\documentclass[prl,twocolumn,preprintnumbers,amsmath,amssymb,showpacs,nofootinbib,floatfix]{revtex4}

\usepackage{graphicx,bm}

\makeatletter
\def\graphicscale{\twocolumn@sw{0.3}{0.4}}
\def\graphicthreescale{\twocolumn@sw{0.3}{0.4}}

\begin{document}

\title{Quantum transitions driven by one-bond defects in quantum Ising
  rings}

\author{Massimo Campostrini,$^1$ 
Andrea Pelissetto,$^2$ and Ettore Vicari$^1$} 

\address{$^1$ Dipartimento di Fisica dell'Universit\`a di Pisa
        and INFN, Largo Pontecorvo 3, I-56127 Pisa, Italy}
\address{$^2$ Dipartimento di Fisica di ``Sapienza" Universit\`a di Roma
        and INFN, Sezione di Roma I, I-00185 Roma, Italy}

\date{\today}

\begin{abstract}

We investigate quantum scaling phenomena driven by lower-dimensional
defects in quantum Ising-like models.  We consider quantum Ising rings
in the presence of a bond defect. In the ordered phase, the system
undergoes a quantum transition driven by the bond defect between a
{\em magnet} phase, in which the gap decreases exponentially with
increasing size, and a {\em kink} phase, in which the gap decreases
instead with a power of the size.  Close to the transition, the system
shows a universal scaling behavior, which we characterize by
computing, either analytically or numerically, scaling functions for
the gap, the susceptibility, and the two-point correlation function.
We discuss the implications of these results for the nonequilibrium
dynamics in the presence of a slowly-varying parallel magnetic field
$h$, when going across the first-order quantum transition at $h=0$.

\end{abstract}


\maketitle



Quantum phase transitions~\cite{Sachdev-book} are phenomena of great
interest in many different branches of physics. They arise in
many-body systems in the presence of competing ground states.  The
driving parameters of the transition are usually bulk quantities, such
as the chemical potential in particle systems, or external magnetic
fields in spin systems. In the presence of first-order transitions,
bulk behavior is particularly sensitive to the boundary conditions or
to localized defects, hence it is possible to induce a quantum
critical transition by changing only the parameters associated with
the defects or the boundaries.

In this paper, we discuss an example of this type of transitions,
considering a quantum Ising ring in a transverse magnetic field with a
bond defect. In the ordered phase, bulk behavior of the low-energy
states depends on the defect coupling. One may have a {\em magnet}
phase, in which the gap decreases exponentially, i.e., $\Delta_L\sim
e^{-cL}$ with increasing the size $L$, or a {\em kink} phase, in which
the lowest states are one-kink states and $\Delta_L \sim 1/L^p$.
Here, we analyze the crossover region between these phases, showing
the emergence of a universal scaling behavior controlled by the defect
coupling. We also analyze the slow nonequilibrium adiabatic dynamics
~\cite{PG-08,PSSV-11} across this transition. We obtain general
time-dependent scaling laws that generalize to the first-order
transition case those that characterize the Kibble-Zurek (KZ)
mechanism at continuous transitions~\cite{Zurek-85,KZ-q,PSSV-11}.

We consider Ising rings of size $L=2\ell+1$ in the presence of a
transverse magnetic field and of one bond defect:
\begin{eqnarray}
H_r = - J \sum_{i=-\ell}^{\ell-1} \sigma^{(1)}_i \sigma^{(1)}_{i+1} 
- g \sum_{i=-\ell}^\ell \sigma^{(3)}_i  
- \zeta \; \sigma_{-\ell}^{(1)} \sigma_\ell^{(1)} ,
\label{hedef}
\end{eqnarray}
where $\sigma_i^{(a)}$ are the Pauli matrices.  We set $J=1$, and
assume $g\ge 0$.  Note that periodic (PBC), open (OBC), and
antiperiodic (ABC) boundary conditions are recovered for $\zeta=1$, 0,
and $-1$, respectively.  The bond defect generally breaks translation
invariance, unless $\zeta=\pm 1$.  A continuous transition occurs at
$g=1$, separating a disordered ($g>1$) from an ordered ($g<1$) phase
\cite{Sachdev-book}.  In the presence of an additional parallel
magnetic field $h$ coupled to $\sigma^{(1)}_i$, a first-order quantum
transition (FOQT) occurs at $h=0$ for any $g < 1$, hence we expect the
defect to be able to change bulk behavior for any $g < 1$. This is the
regime we shall consider below.

Analytic and accurate numerical results can be obtained by exploiting
the equivalent quadratic fermionic Hamiltonian which is obtained by a
Jordan-Wigner transformation~\cite{LSM-61,footnote-n}. We analyze the
dependence of low-energy properties on the defect parameter $\zeta$
~\cite{inprep}.  In particular, we consider the energy differences
\begin{equation}
\Delta_{L,n}\equiv E_n-E_0,\qquad \Delta_L \equiv \Delta_{L,1},
\label{deltaldef}
\end{equation}
where $E_0$ is the energy of the ground state, and $E_{n\ge 1}$ are
the (ordered) energies of the excited levels.  The magnetization
$\langle \sigma_x^{(1)} \rangle$ vanishes due to the global ${\mathbb
  Z}_2$ symmetry. Thus, we use the two-point correlation function
$G(x,y) \equiv \langle \sigma_x^{(1)} \sigma_y^{(1)} \rangle$ to
characterize the magnetic properties of the ground state.

For $g < 1$, we should distinguish a {\em magnet} phase ($\zeta > -1$)
and a {\em kink} phase ($\zeta \le -1$).  The lowest states of the
magnet phase are superpositions of states with opposite nonzero
magnetization $\langle\pm |\sigma_x^{(1)}|\pm\rangle = \pm m_0$
(neglecting local effects at the defect), where~\cite{Pfeuty-70}
$m_0=(1-g^2)^{1/8}$. For a finite chain, tunneling effects between the
states $| + \rangle$ and $| -\rangle$ lift the degeneracy, giving rise
to an exponentially small gap $\Delta_L$ \cite{ZJ-86,BC-87}.  For
example,~\cite{Pfeuty-70} $\Delta_L \approx 2 (1-g^2) g^L$ for
$\zeta=0$ (OBC).  An analytic calculation gives
\begin{equation}
\Delta_L \approx {8g\over 1-g} w^2 e^{-wL},
\qquad  w={1-g\over g} \;(1+\zeta),
\label{defde}
\end{equation}
for $\zeta\to -1^+$.  The large-$L$ two-point function is trivial,
\begin{equation}
G_r(x_1,x_2) \equiv {G(x_1,x_2)\over m_0^2} \to 1
\label{gcmagnet}
\end{equation}
for $x_1\neq x_2$, keeping $X_i\equiv x_i/\ell$ fixed (but $X_i\neq
\pm 1$).

The low-energy behavior drastically changes for $\zeta\le -1$, in
which the low-energy states are one-kink states (made of a
nearest-neighbor pair of antiparallel spins), which behave as
one-particle states with $O(L^{-1})$ momenta~\cite{Sachdev-book}. In
particular, for $\zeta=-1$ (ABC) we have
\begin{equation}
\Delta_L = {g\over 1-g} \, {\pi^2\over L^2} + O(L^{-4}).
\label{deltaabc}
\end{equation}
The first two excited states are degenerate, thus
$\Delta_{L,2}=\Delta_{L,1}\equiv \Delta_L$.  For $\zeta<-1$, the
ground state and the first excited state are superpositions with
definite parity of the lowest kink $|\downarrow\uparrow\rangle$ and
antikink $|\uparrow\downarrow\rangle$ states.  The gap scales as
\cite{CJ-87,BC-87} $L^{-3}$; we obtain explicitly
\begin{eqnarray}
\Delta_L = 
{8 \zeta g^2 \over (1-\zeta^2) (1-g)^2}\,{\pi^2\over L^3} + O(L^{-4}).
\label{deltallm3}
\end{eqnarray}
On the other hand, $\Delta_{L,n}$ for $n\ge 2$ behaves as $L^{-2}$,
e.g.,
\begin{equation} 
\Delta_{L,2} = { 3 g \over (1-g)}\,{\pi^2\over L^2} 
+ {6(1-\zeta)g^2\over (1+\zeta)(1-g)^2} \,{\pi^2\over L^3} + 
O(L^{-4}).
\label{deltallm32}
\end{equation}
The two-point function $G(x,y)$ can be perturbatively computed for
small $g$, obtaining the asymptotic large-$L$ behaviors
\begin{eqnarray}
&&G(x_1,x_2) =  1 - |X_1-X_2| \qquad {\rm for}\;\; \zeta=-1,
\label{gcxzm1}\\
&&G(x_1,x_2) = 1 - |X_1-X_2| - 
{|\sin(\pi X_1)-\sin(\pi X_2)| \over \pi}\qquad
\label{gczetamm1}
\end{eqnarray}
for $\zeta<-1$, where $X_i\equiv x_i/\ell$.  We conjecture (and verify
numerically) that the above formulas can be straightforwardly
generalized to the whole ordered phase $g<1$ by simply introducing a
multiplicative renormalization, i.e., by replacing $G$ with $G_r\equiv
G/m_0^2$.

\begin{figure}[tbp]
\includegraphics*[scale=\graphicscale]{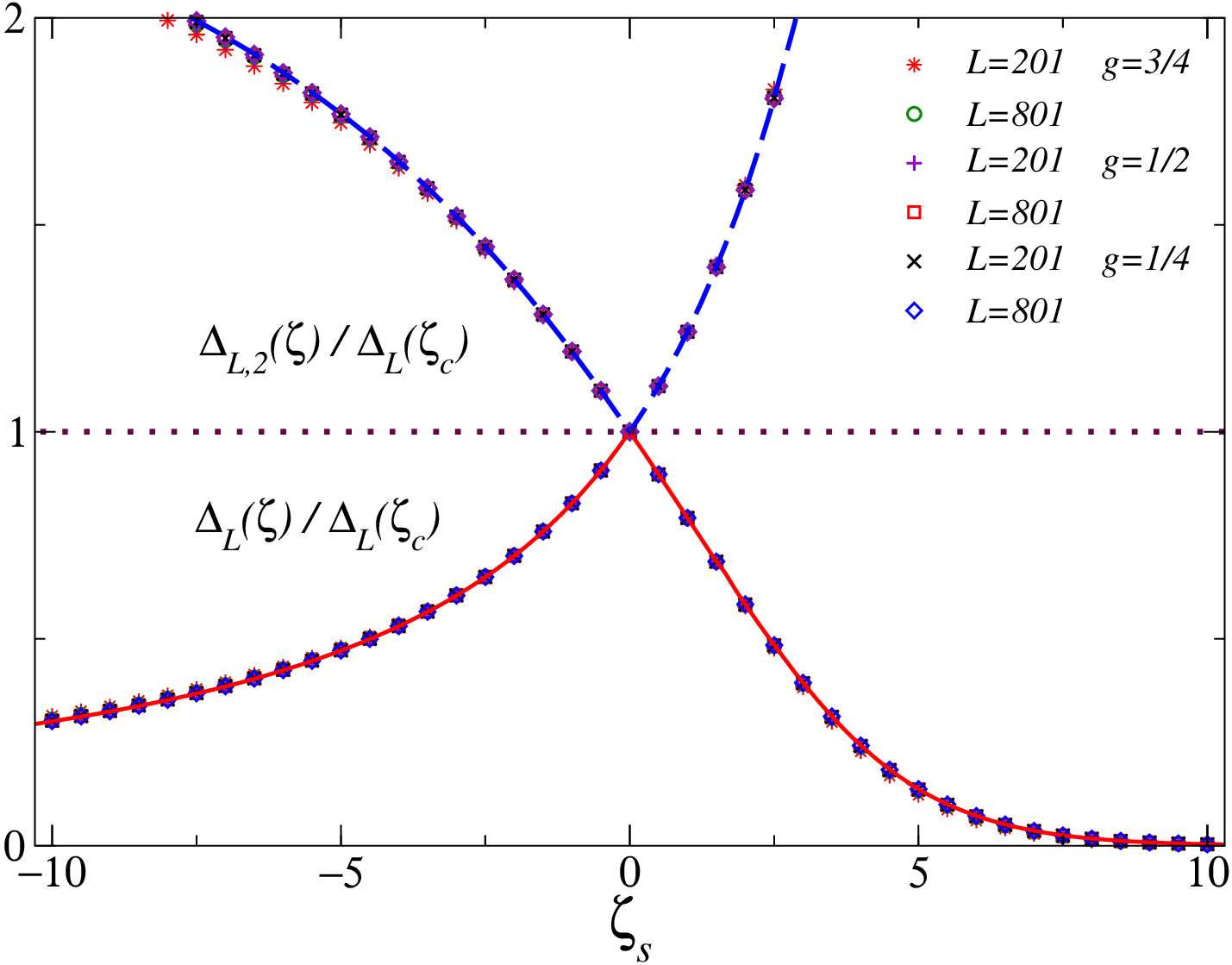}
\caption{(Color online) We show the scaling functions $D_n(\zeta_s)$,
  cf. Eq.~(\ref{deltasca}), and numerical data for the ratio
  $\Delta_{L,n}(\zeta)/\Delta_{L}(\zeta_c)$, for $n=1$ (bottom) and
  $n=2$ (top) separated by the dotted line.  Numerical data clearly
  approach the $g$-independent scaling curves $D_n(\zeta_s)$
  (differences are hardly visible).  }
\label{dezetac}
\end{figure}

These results suggest that $\zeta_c=-1$ is a critical point,
separating the magnet and kink phases.  We now show that around
$\zeta_c$ the system develops a universal scaling behavior.  We
analytically compute (and verify numerically) the asymptotic behavior
of $\Delta_{L,n}(\zeta)$, obtaining the scaling behavior
\begin{eqnarray}
&&\Delta_{L,n}(\zeta) \approx \Delta_{L}(\zeta_c)\, D_{n}(\zeta_s),
\quad \label{deltasca}\\
&&\zeta_s = {1-g\over g} (\zeta - \zeta_c)\,L, \qquad \zeta_c=-1,
\label{zetasdef}
\end{eqnarray}
for $L\to \infty$ keeping the scaling variable $\zeta_s$ fixed. The
scaling functions $D_1$ and $D_2$ are shown in
Fig.~\ref{dezetac}~\cite{footnote-asybehd}. The cusp-like behavior at
$\zeta_s=0$ is the consequence of the crossing of the first two
excited states at $\zeta=-1$.  Generally, the scaling functions are
universal apart from normalizations of their arguments. In this case,
the normalization of $\zeta_s$ is chosen so that the scaling curves
for different values of $g$ are identical.  Notice that, once the
normalization is fixed by using one observable, universality should
hold for any other observable. Numerical results for the energy
differences $\Delta_{L,n}$ are shown in Fig.~\ref{dezetac}: they
confirm the scaling behavior (\ref{deltasca}).  The asymptotic
large-$L$ behavior is generally approached with corrections of order
$L^{-1}$.

\begin{figure}[tbp]
\includegraphics*[scale=\graphicscale]{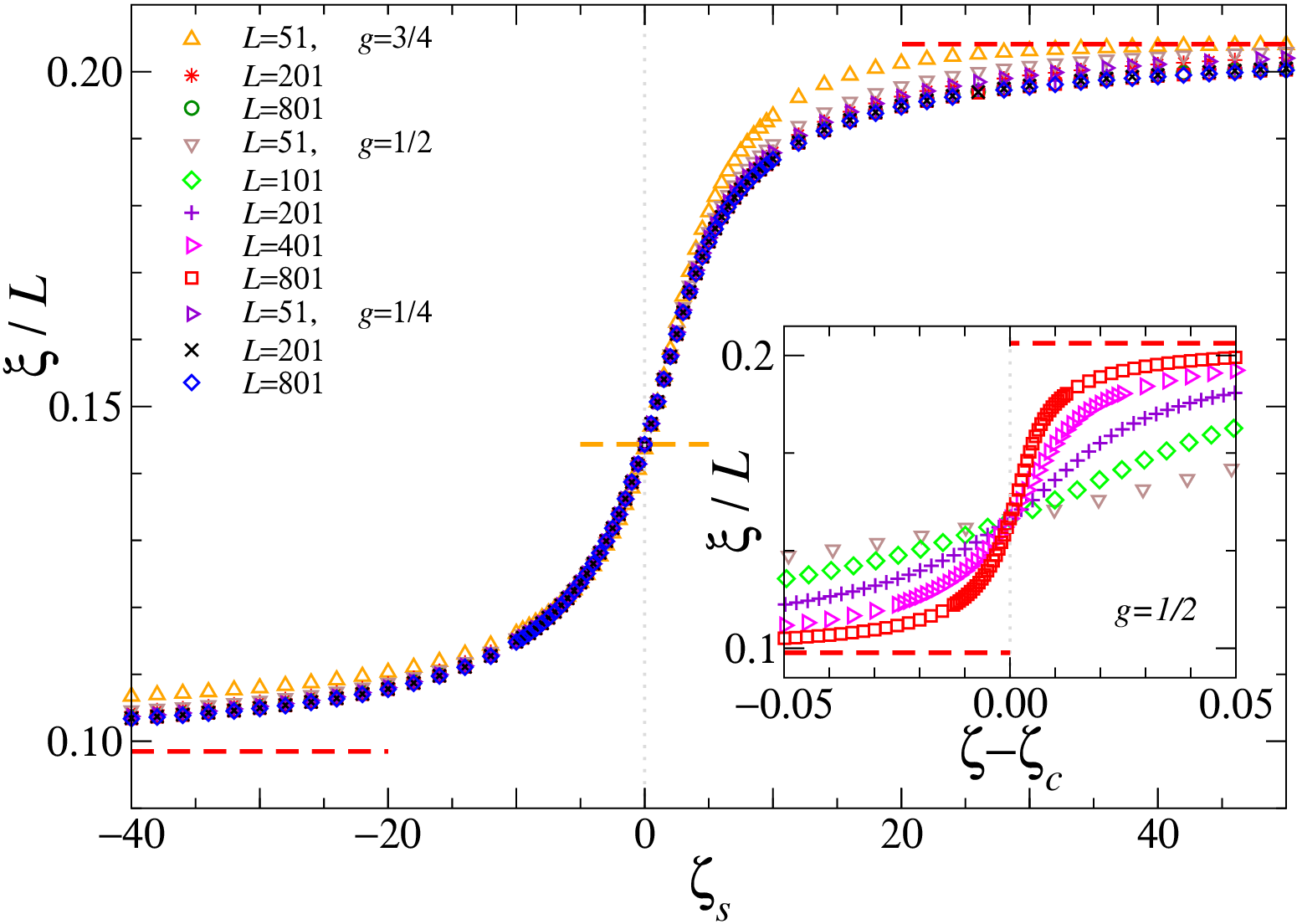}
\caption{(Color online) Estimates of the ratio $\xi/L$, supporting the
  scaling ansatz (\ref{xiolsca}).  Scaling corrections are only
  visible for $L\lesssim 100$.  The dashed lines indicate the values
  of $f_\xi(\zeta_s)$ for $\zeta_s\to \pm \infty$ and $\zeta_s=0$,
  obtained by matching the scaling ansatz with the behaviors in the
  different phases: $f_\xi(\infty) = 1/\sqrt{24}$ from
  Eq.~(\ref{gcmagnet}), $f_\xi(-\infty)\approx 0.098491$ from
  Eq.~(\ref{gczetamm1}), and $f_\xi(0)=1/\sqrt{48}$ from
  Eq.~(\ref{gcxzm1}). The inset shows the crossing point of data for
  different $L$ implied by Eq.~(\ref{xiolsca}).}
\label{xisca}
\end{figure}

Other observables satisfy analogous scaling relations.  The two-point
function is expected to behave as
\begin{equation}
G(x_1,x_2;\zeta) \approx m_0^2 
\; {\cal G}(X_1,X_2;\zeta_s),\qquad X_i=x/\ell,\quad
\label{gscal}
\end{equation}
where $m_0=(1-g^2)^{1/8}$. This scaling ansatz can be checked by
considering the zero-momentum quantities
\begin{equation}
\chi = \sum_x G(0,x), \qquad
\xi^2 = {1\over 2 \chi} \sum_x x^2 G(0,x)\label{chixidef}
\end{equation}
($\xi$ is the second-moment length scale). Eq.~(\ref{gscal}) implies
\begin{equation}
\chi/L \approx m_0^2 f_\chi(\zeta_s),\qquad
\xi/L \approx f_\xi(\zeta_s).  
\label{xiolsca}
\end{equation}
Numerical data confirm them, see Fig.~\ref{xisca}.  In the language of
renormalization-group (RG) theory, the defect coupling $\zeta$ plays
the role of a relevant parameter at the magnet-kink transition, with
RG dimension $y_\zeta=1$.

\begin{figure}[tbp]
\includegraphics*[scale=\graphicscale]{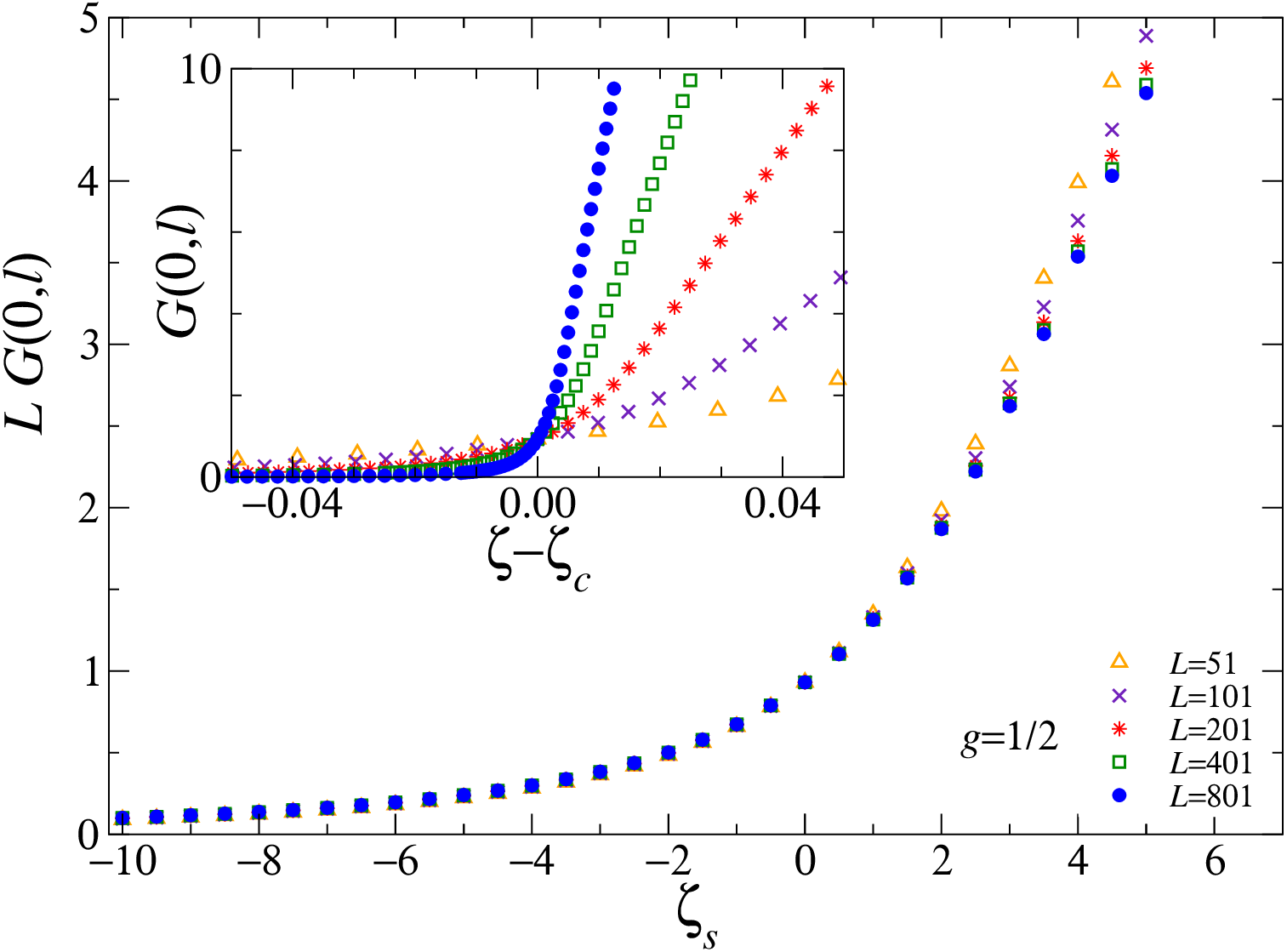}
\caption{(Color online) Scaling behavior of the center-defect
  correlation $G(0,\ell)$. The data of $LG(0,\ell)$ approach a scaling
  function of $\zeta_s$.  The inset shows $G(0,\ell)$ as a function of
  $\zeta - \zeta_c$.  }
\label{bzetar}
\end{figure}

An interesting question is whether there is a quantity playing the
role of order parameter for the magnet-kink transition.  This is
provided by the center-defect correlation $b = \lim_{L\to\infty}
G(0,\ell)$.  Indeed, $b > 0$ for $\zeta>\zeta_c$ and $b = 0$ for
$\zeta\le \zeta_c$.  Moreover, we observe the scaling
behavior~\cite{footnotegll} $G(0,\ell) \sim L^{-1} f_b(\zeta_s)$, with
$f_b(-\infty)=0$ and $f_b(\infty)= \infty$,
see Fig.~\ref{bzetar}.

An analogous magnet-to-kink quantum transition can be observed in the
Ising chain by appropriately tuning a magnetic field $\eta$, coupled
to $\sigma^{(1)}$, localized at the boundaries. Explicitly, we
consider (we assume $\eta\ge 0$)
\begin{eqnarray}
H_c = - \sum_{i=-\ell}^{\ell-1} \sigma^{(1)}_i \sigma^{(1)}_{i+1} 
- g \sum_{i=-\ell}^\ell \sigma^{(3)}_i  
- \eta \; (\sigma_{-\ell}^{(1)} - \sigma_\ell^{(1)}). \quad
\label{hedef2}
\end{eqnarray}
The analytic computation of the low-energy spectrum identifies a
particular value of the boundary field, $\eta_c = \sqrt{1-g}$,
separating the magnet and kink phases.  In the magnet phase $\eta <
\eta_c$ we have
\begin{eqnarray}
\Delta_L = { 2 gs \sqrt{s^2-1} \over s-g} \; s^{-L} + O(s^{-2L}),\quad
s = {1-\eta^2\over g}.
\label{deltalf}
\end{eqnarray}
For $\eta \ge \eta_c$ we have instead 
\begin{eqnarray}
\Delta_L = c(\eta) {g\over 1-g} \, {\pi^2\over L^2} + O(L^{-3}) 
\label{deltalk}
\end{eqnarray}
with $c=1$ for $\eta=\eta_c$ and $c=3$ for any $\eta>\eta_c$.  The
similar nature of the coexisting phases suggests that their asymptotic
large-$L$ scaling behavior for $\zeta \approx \zeta_c$ and $\eta
\approx \eta_c$ is the same, i.e., the two transitions belong to the
same universality class.  This is confirmed by analytic and numerical
computations, although the comparison of the results is not
straightforward, as the Ising chain (\ref{hedef2}) breaks the
${\mathbb Z}_2$ symmetry, which is instead preserved by the Ising ring
(\ref{hedef}).  For example, the gap satisfies scaling relations
analogous to Eq.~(\ref{deltasca}). Explicitly,
\begin{equation}
\Delta_{L,n}(\eta) = \Delta_L(\eta_c)  E_n(\eta_s), \quad \eta_s =
      {2\sqrt{1-g}\over g}(\eta_c-\eta)L,
\label{deltaeta}
\end{equation}
with \cite{footnote_rescaling} $E_n(x) = D_{2n - 1}(x)$ for $x \ge 0$
and $E_n(x) = D_{2n}(x)$ for $x \le 0$.  The reason of the peculiar
mapping is related to the different behavior under ${\mathbb Z}_2$ of
the two models.  Consider for instance the kink phase.  While in the
Ising ring the lowest states are superpositions of kink and antikink
configurations, in the case of model (\ref{hedef2}) the parity
symmetry is broken by the boundary fields, thus only kink states are
left.  Hence, model (\ref{hedef2}) has only half of the states of the
Ising ring. Moreover, no degeneracy occurs at $\eta = \eta_c$ so that
levels must be smooth at the transition point, thereby explaining why
the mapping between the levels must change at the transition point (in
the ring case, cusps occur at the transition).

It is worth noting that, at the critical value $g=1$, corresponding to
the order-disorder continuous transition, bulk behavior is independent
of the boundary conditions or of the presence of defects, hence the
magnet-to-kink transition only occurs for $g$ strictly less than 1.
For instance, the gap at $g=1$ behaves as $\Delta_L\sim L^{-1}$ for
any $\zeta$ or $\eta$. Of course, the prefactor depends on the
boundary conditions; see, e.g., the known results for PBC, OBC, and
ABC, summarized in Ref.~\cite{CPV-14}.

It is interesting to reinterpret our results in the equivalent
fermionic picture of models (\ref{hedef}) and (\ref{hedef2}).  In the
magnet phase, i.e., for $\zeta>\zeta_c$ and $\eta<\eta_c$,
respectively, the lowest eigenstates are superpositions of Majorana
fermionic states localized at the boundaries or on the
defect~\cite{Kitaev-01,Alicea-12}.  In finite systems, their overlap
does not vanish, giving rise to the splitting $\Delta \sim
e^{-L/l_0}$.  The coherence length $l_0$ diverges at the
kink-to-magnet transitions as $l_0^{-1}\sim |\ln s|\sim \eta_c-\eta$
and $l_0^{-1}\sim \zeta-\zeta_c$ in the two models, a behavior
analogous to that observed at the order-disorder transition $g\to 1^-$
where $l_0^{-1}\sim |\ln g|$.

In conclusion, we have shown that quantum transitions can be induced
by tuning the boundary conditions or by changing lower-dimensional
defect parameters, when the system is at a FOQT. We have explicitly
discussed this behavior in the case of quantum Ising rings in a
transverse field. If $g<1$, this model shows a magnet and a kink
phase, separated by a quantum transition point. In its neighborhood,
we can define general scaling laws, that are analogous to those that
hold at continuous transitions.  The same scaling behavior is also
observed in the XY quantum ring in which one adds additional bond
couplings $\sigma^{(2)}_i \sigma^{(2)}_{i+1}$ \cite{footnoteXY}, and
in the quantum Ising chain with opposite magnetic fields at the
boundaries.  The universal scaling behavior is essentially the same
and is uniquely determined by the structure of the low-energy behavior
in the two phases. We have characterized the scaling variable and
computed the scaling functions of different observables. These scaling
behaviors can be straightforwardly extended to allow for a nonzero
temperature $T$, by considering a further dependence on the scaling
variable $TL^z=TL^2$.  Even though we have discussed the issue in one
dimension, we expect the same type of behavior in quantum
$d$-dimensional Ising models defined in $L^{d-1}\times M$ boxes with
$L\gg M$, in the presence of a $(d-1)$-dimensional surface of defects
or of opposite magnetic fields on the $L^{d-1}$ boundaries.

The size dependence of the low-energy spectrum is relevant for the
understanding of the nonequilibrium unitary dynamics, as it determines
the conditions for a nearly adiabatic quantum
dynamics~\cite{PG-08,PSSV-11}.  Significantly different behaviors are
expected in the magnet and kink phases when we add a time-dependent
parallel magnetic term,
\begin{equation}
H_t = H - h(t/t_0)\sum_i \sigma_i^{(1)},\qquad h(u)=h_0 u,
\label{hh}
\end{equation}
where $t_0$ is the time scale of the time dependence.  Adiabatic
evolutions across the FOQT ($h(0)=0$) require very different time
scales $t_0$. In the magnet phase we must have $t_0 \gtrsim
e^{2L/l_0}$, while in the kink phase $t_0 \gtrsim L^4$ at $\zeta_c$
($\eta_c$) and $t_0 \gtrsim L^6$ for $\zeta<\zeta_c$ [$t_0\gtrsim L^4$
  for $\eta>\eta_c$ in model (\ref{hedef2})].  The dynamics in the
magnet phase is essentially equivalent to that of a two-level
Landau-Zener model with an exponentially small
gap~\cite{CNPV-14,DMSGG-97}. At $\zeta_c$ for model (\ref{hedef}) and
$\eta\ge \eta_c$ for model (\ref{hedef2}), dynamics becomes similar to
that at a continuous transition with a dynamic exponent $z=2$, since
there is a tower of excited states with
$\Delta_{L,n}=O(L^{-2})$. Model (\ref{hedef}) for $\zeta<\zeta_c$
shows again a low-energy dynamics dominated by the two lowest states
with a gap of order $L^{-3}$.  The above results suggest that
nonequilibrium scaling laws, such as those describing the KZ mechanism
at continuous transitions~\cite{Zurek-85,KZ-q,PSSV-11} (when systems
are ramped across a continuous transition at a finite rate), also hold
at FOQTs, when the time-dependent $h$ crosses the value $h=0$.  The
slow nonequilibrium dynamics of the Hamiltonian (\ref{hh}) across the
FOQT $h=0$, and in particular the interplay among $t$, $t_0$, $\zeta$
(or $\eta$) and $L$, thus $\Delta_L$, can be described by a scaling
theory which extends the equilibrium scaling behaviors (\ref{gscal})
and (\ref{xiolsca}).  For this purpose, we use scaling arguments
similar to those employed to describe the KZ problem~\cite{KZ-q}. If
$y_h = d + z$ ($d$ is the space dimension, $d=1$ in our example) is
the effective RG dimension of the parallel field $h$ ($z = 2$ and $y_h
= 3$ in the Ising chain \cite{CNPV-15}), we can define an effective
length scale $\xi_h$ associated with $h$ that scales as $\xi_h \sim
h^{-1/y_h}$.  The dynamical scaling laws for a system of size $L$ can
be heuristically derived by substituting $h$ with the time-dependent
$h(t/t_0)$ into the scaling combinations $t \xi_h^{-z}=t h^{z/y_h}$
and $\xi_h/L$.  This gives $t(t/t_0)^{z/y_h} = (t/\tau)^{1+z/y_h}$ and
$L/\xi_h = L (t/t_0)^{1/y_h}= (t/\tau)^{1/y_h} L/\tau^{1/z}$, with
$\tau=t_0^{z/(z+y_h)}=t_0^{2/5}$~\cite{footnoteKZ}. These
considerations lead us to conjecture the scaling behavior for the
Ising chain
\begin{eqnarray}
&&\langle \sigma_x^{(1)} \rangle \approx m_0 \,f_m\left(x/L, \,t/\tau, 
\, \tau/L^2, \,\zeta_s\right),
\label{mxt}\\
&&\langle \sigma_{x_1}^{(1)} \sigma_{x_2}^{(1)} \rangle
 \approx m_0^2 \,f_g\left(x_i/L, \, t/\tau, \, \tau/L^2 , \,\zeta_s\right),
\label{KZsca}
\end{eqnarray}
with $\zeta_s$ given by Eq.~(\ref{zetasdef}). Analogous expressions
apply to other observables and to the model (\ref{hedef2}). Of course,
the above nonequilibrium scaling theory should be further investigated, to
get a thorough understanding of KZ-like phenomena at FOQTs.

These issues may also be relevant in the context of quantum computing.
Adiabatic algorithms rely on sufficiently large gaps during the
variation of the model parameters bringing to the ground state of the
desired Hamiltonian~\cite{FGGS-00,FGGLLP-01,Bloch-08}.  Thus, FOQTs,
at which the gap is exponentially small, represent a hard
problem~\cite{AC-09,YKS-10,LMSS-12}.  As a simple paradigmatic case we
may consider the time-dependent Hamiltonian (\ref{hh}) for $g<1$ and
$\zeta=1$.  Let us assume that we want to adiabatically move from the
ground state with $h=-h_0$ at time $t=-T_a$ to the ground state with
$h=h_0$ at $t=T_a$.  This requires an exponentially large time scale,
i.e. $T_a \gtrsim e^{2L/l_0}$.  Our results for the $\zeta$-dependence
of the low-energy properties suggest a way to overcome this hard
problem. Indeed, instead of changing directly $h$, one could proceed
as follows.  First, one adiabatically changes the system varying the
bond defect from $\zeta=1$ to $\zeta\lesssim \zeta_c$, which
corresponds to adding a further single-bond Hamiltonian term $H_\zeta
= - \zeta(t/T_\zeta)\,\sigma_{-\ell}^{(1)}\sigma_{\ell}^{(1)}$.  In
the presence of a finite parallel magnetic field $h=-h_0$, the gap is
finite. Then, one adiabatically changes $h$, according to
Eq.~(\ref{hh}), which now requires a time scale $T_a\gtrsim L^4$ or
$L^6$ to adiabatically go from $-h_0$ to $h_0$. Finally, $\zeta$ is
increased again to $\zeta = 1$, obtaining the ground state of the
original problem in a total time that scales with a power, and not
with the exponential, of the size.  Therefore, by taking advantage of
particular sensitivity of FOQTs to defects or boundary perturbations,
one may overcome the problem of an exponentially slow dynamics, which
occurs at FOQTs within the more standard approaches.


\begin{thebibliography}{99}
\bibitem{footnote-n} Numerical computations are similar to those
  reported in the literature (see, e.g., Ref.~\cite{YR-96}), but with
  an additional complication for $\zeta\le 0$: the ground state
  corresponds to the first excited state of the fermionic model.
\bibitem{inprep}
We present several analytic results.
Derivations will be reported elsewhere.
\bibitem{footnotegll} Notice that the scaling behavior of $G(0,\ell)$
  cannot be derived from Eq.~(\ref{gscal}).
\bibitem{footnote_rescaling} In general one would also expect an
  argument rescaling; we have defined $\eta_s$ in such a way to make
  the rescaling factor equal to 1 for all values of $g$.
\bibitem{footnoteXY} We numerically checked that the universality of
  the scaling phenomena driven by one defect extends to the XY ring,
  $H_{XY} = - \sum_{i=-\ell}^{\ell-1} {\cal H}_{i,i+1} - g
  \sum_{i=-\ell}^\ell \sigma^{(3)}_i - \zeta \,{\cal H}_{-\ell,\ell}$
  where ${\cal H}_{i,j} = {1+\gamma\over 2} \sigma^{(1)}_i
  \sigma^{(1)}_{j} + {1-\gamma\over 2} \sigma^{(2)}_i
  \sigma^{(2)}_{j}$, when varying $\gamma$ around $\gamma=1$ (e.g.,
  for $g=1/2$ and $\gamma=3/4$).  Different behaviors are instead
  observed at small values of $\gamma$ (e.g., $g=1/2$ and
  $\gamma=1/2$).  This is related to the fact that the low-energy
  spectrum qualitatively changes when $|\gamma|<\sqrt{1-g}$. If
  $\gamma$ is small, the kink phase no longer exists (the ground-state
  properties of the magnet phase for $\zeta>\zeta_c$ remains stable).
  For $\zeta=-1$ (ABC), the gap behaves as $\Delta_L \approx a/L^2$
  with $a = {(g-1+\gamma^2)\pi^2/(1 - g)}$, which vanishes when
  $\gamma-\sqrt{1-g}\to 0^+$.  For $\gamma>\sqrt{1-g}$ the scaling
  variable (\ref{zetasdef}) is replaced by $\zeta_s={(1-g)(\zeta+1)L/
    (g-1+\gamma^2)}$.
\bibitem{footnoteKZ} The same arguments can be applied to the KZ
  problem driven by the parallel field $h(t)=t/t_0$ at the
  order-disorder continuous transition ($g=1$), which belongs to the
  2D Ising universality class with $z=1$ and $y_h=2-\eta/2=15/8$. They
  give $\tau = t_0^{z/(z+y_h)}=t_0^{8/23}$. This provides the scaling
  behavior for the KZ problem driven by the parallel magnetic field.
  The scaling of the KZ problem when varying the transverse field $g$
  across $g=1$ is obtained by replacing $y_h$ with
  $y_g=1/\nu=1$~\cite{KCH-12}.
\end{thebibliography}
\end{document}